\documentclass[12pt,preprint]{aastex}
\slugcomment{submitted to ApJ}

\shorttitle{Core He-burning and White Dwarf composition}
\shortauthors{Straniero, Dom\'\i nguez, Imbriani \& Piersanti}
\begin{document}
\title{The chemical composition of White Dwarfs as a test of convective
efficiency during core He-burning}

\author {Oscar Straniero}
\affil{Osservatorio Astronomico di Collurania, 64100 Teramo, Italy \\ straniero@te.astro.it}
\author{Inmaculada Dom\'\i nguez}
\affil{Dpto. de F\'{\i}sica Te\'orica y del Cosmos, Universidad de Granada,
18071 Granada, Spain\\ inma@ugr.es}
\author {Gianluca Imbriani}
\affil{Osservatorio Astronomico di Collurania, 
64100 Teramo, and INFN Napoli, Italy\\ gianluca.imbriani@na.infn.it}
\and
\author {Luciano Piersanti}
\affil{Osservatorio Astronomico di Collurania, 64100 Teramo, Italy \\ piersanti@te.astro.it}
 
\begin{abstract}

Pulsating white dwarfs provide constraints to the evolution of 
progenitor stars. We revise 
He-burning stellar models, with particular attention
to core convection and to its connection with the nuclear
reactions powering energy generation and chemical evolution. 
Theoretical results are compared to the available
measurements for the variable white dwarf GD 358, which indicate a
rather large abundance of central oxygen (Metcalfe, Winget \& Charbonneau 2001).
 We show that the attempt to constrain the 
relevant nuclear reaction rate by means of the white dwarf composition is faced
with a large degree of uncertainty related to evaluating the efficiency of
convection-induced mixing.
By combining the uncertainty of the convection theory with the error on the relevant reaction rate
we derive that the present theoretical prediction for the central oxygen mass fraction
in white dwarfs varies between 0.3 and 0.9.
Unlike previous claims, we find that models taking into account 
semiconvection and a moderate $^{12}$C$(\alpha,\gamma)^{16}$O reaction 
rate are able to account for a high
central oxygen abundance. The rate of the $^{12}$C$(\alpha,\gamma)^{16}$O used in these models  
agrees with the one recently obtained in laboratory experiments (Kunz et al. 2002). 
On the other hand, when semiconvection is inhibited, as in the case of 
classical models (bare Schwarzschild criterion)
or in models with mechanical overshoot, an extremely high rate of the 
$^{12}$C$(\alpha,\gamma)^{16}$O reaction is needed to account 
for a large oxygen production. 
Finally, we  show that the apparent discrepancy between our result and those reported in previous studies  
depends on the method used to avoid the convective runaways (the so called breathing pulses), which are
usually encountered in modeling late stage of core He-burning phase.

\end{abstract} 

\keywords {white dwarfs - stars:evolution - star:interiors -  nuclear reactions - convection}

\section {Introduction}
White dwarfs (WDs) are dead stars, which supply the energy irradiated from the surface 
by consuming their thermal reservoir. Since the thermal content of a WD depends on its 
chemical composition (see Van Horn 1971), evaluation of the cooling time scale 
requires a good knowledge
of the processes that modify the original composition of the progenitor. In addition,
some observed
features of type Ia Supernova outburst depend on the  
detailed internal
composition of the exploding WD 
(Dom\'\i nguez, H\"oflich \& Straniero, 2001). According to the theory of
stellar evolution, the majority of WDs are post Asymptotic Giant Branch (AGB)
stars (Paczi\'nsky 1970a).
In such a case, they would consist of the primary ashes of He-burning, 
essentially  carbon and oxygen:
C is initially produced by the $3\alpha$ reactions and, subsequently,
O is synthesized via the 
$^{12}$C$(\alpha,\gamma)^{16}$O. 

During their long
cooling time, white dwarfs cross the instability strip and undergo stable pulsations. 
These pulsations can be observed as variations in brightness, which could be used to 
provide constraints on the internal structure
of these condensed objects. This is an important tool to test the reliability of
progenitor models.
Metcalfe, Nather and Winget (2000) 
analysed seismic data of the variable GD 358, a DB type white dwarf,  
and estimated the WD mass, the effective temperature and the mass of the thin 
helium-rich envelope. 
This method was recently extended to provide a description of
the internal chemical profile (Metcalfe, Winget, Charbonneau, 2001; 
see also Metcalfe, Salaris \& Winget, 2002).
In particular, a  
large oxygen abundance was found in the innermost region, namely $0.84 \pm 0.03$.
By comparing this chemical profile with
those predicted by theoretical models of WD progenitors,
it was concluded that the reaction rate of the  
$^{12}$C$(\alpha,\gamma)^{16}$O is definitely larger than those 
recently derived in laboratory experiments (Buchman et al. 1996, Kunz et al. 2002). 

We believe that these constraints on the rate of the $^{12}$C$(\alpha,\gamma)^{16}$O 
reaction are model dependent. We will show how the predicted WD composition changes when the scheme
for the convective mixing is changed. The connection between nuclear reaction and
convection and their influence on the final amounts of C and O in the He-exhausted 
cores of stars with a mass between 0.8 and 25 M$_\odot$  have been recently revised 
by Imbriani et al. (2001). In particular, they showed that oxygen production 
via $^{12}$C$(\alpha,\gamma)^{16}$O may be substantially
increased by extending the central mixing during the late stage of the 
core He-burning phase, when the central He mass fraction falls below about 0.1.

Unfortunately, in spite of the many theoretical works published over the last
three decades, the
physics that determines the extent of the convective region within the He core is still
poorly known. The theoretical calculations available so far leave various scenarios open.
Classical models,
those based on a bare Schwarzschild criterion (as early presented by Iben \& Rood, 1970),
are still 
calculated and widely used in many studies (e.g. Umeda et al. 1999, Althaus et al. 2002). 
Nonetheless, models that include some algorithm to handle the discontinuity of the opacity
that forms at the external border of the convective core, as a consequence of the conversion
of He into C (and O), should be considered as more reliable 
(Paczi\'nsky 1970b, Castellani, Giannone \& Renzini 1971a and 1971b, Demarque \& Mengel 1972,
Robertson \& Faulkner 1972, Sweigart \& Demarque 1972,
Sweigart \& Gross 1976, Castellani et al. 1985, Iben 1986, Sweigart  1990, 
Lattanzio 1991, Dorman \& Rood 1993, 
Dom\'\i nguez et al. 1999, and references therein). We remind that this phenomenon naturally leads to
the growth of the convective core (the so called {\it induced overshoot})
 and to the formation of a semiconvective layer outside
the fully convective region.
Alternative models assume that a mechanical overshoot (i.e. due to 
the inertia of the material accelerated by the buoyancy forces) takes place at the boundary of the 
convective region (Saslaw \& Schwarzschild 1965, 
Shaviv \& Salpeter 1973, Maeder 1975, Bertelli, Bressan \& Chiosi 1985, Bertelli et al. 1990, 
Shaller et al. 1992, Bressan et al. 1993, 
Herwig et al. 1997, Girardi et al. 2000, and references therein). 
The existence of this phenomenon can not be questioned 
in the framework of a reliable physical scenario. However, the quantitative relevance of the
mechanical overshoot is a matter of a hard debate (see e.g. Renzini 1987). The most recent
attempts to calibrate the extension of the convective overshoot for centrally H-burning stars
conclude that it should be confined within 0.2-0.3 $H_p$,
where $H_p$ is the pressure scale height (Stothers \& Chin 1992, Bressan et al. 1993, 
Demarque, Sarajedini \& Guo 1994,
Mermilliod, Huestamendia \& del Rio 1994, Schr\"oder, Pols \& Eggleton 1997, Testa et al. 1999). 
Concerning the core He-burning phase, a moderate mechanical overshoot   
mimics the effect of the induced overshoot.
On the contrary, a large mechanical overshoot (say 1 $H_P$ or larger) would cancel out 
the semiconvective zone and major changes in the theoretical scenario are
expected.

Since the theoretical uncertainty does not provide a satisfactory answer 
to the problem of stellar convection, 
some observational constraints have been
investigated.
In particular, the
luminosity function of bright Globular Cluster stars could be used to 
provide some indications on the correct mixing scheme.
The larger the mixing during the core He-burning, the less He fuel
will be left for the subsequent AGB evolutionary phase. Thus, 
the ratio $R_2$ (number of stars observed in the AGB/number of stars observed in the Horizontal Branch)
may be used to constrain the mixing efficiency (see, for example, the discussion
in Renzini \& Fusi Pecci 1988).       
The measured values of $R_2$ in Globular Clusters ($0.15\pm0.01$: see Buzzoni et al 1983, 
Buonanno et al. 1985) support semiconvective models.
An alternative method, based on Red Giant Branch stars,
has been proposed by Caputo et al. (1989); their conclusions agree with those based on the 
$R_2$ parameter.

The knowledge of the internal composition
of WDs, provides an independent method to constrain the physics controlling  
chemical variations in the core of He-burning stars.
In this context, the result obtained by Metcalfe and co-workers 
appears in contrast with the indications 
derived from the Globular Cluster luminosity functions. In fact, they showed 
that semiconvective models can not reproduce the large value of the oxygen measured in GD 358.
As a possible solution of this problem, they propose a substantial enhancement of the 
rate of the $^{12}$C$(\alpha,\gamma)^{16}$O reaction with respect to the experimental values.
In this paper we review various convective schemes and we discuss their 
impact on the predicted core composition.
Contradicting previous claims, we will show that
fully semiconvective models can account for a relatively
high value of the central oxygen, even if a moderate
rate of the $^{12}$C$(\alpha,\gamma)^{16}$O reaction is used, in agreement with the
most recent laboratory measurements.

\section{Five models for the core He-burning phase} 
In this section we revise the theoretical expectations for core He-burning models. 
The stellar structure at the beginning of the core He burning phase has been obtained by evolving
a 3 M$_\odot$ model with solar composition (Z=0.02, Y=0.28), since the pre-Main Sequence. 
Then, five He-burning models have been calculated
under different assumptions for the convective scheme. We have selected 
the mixing algorithms by searching in the recent literature for the models more
representative of the theoretical scenarios typically adopted for He-burning stars.   
The choice for the initial stellar parameters is adequate for the
progenitor of a disk WD with mass 
of about $\sim 0.65$ M$_\odot$, 
which is the mass estimated by Metcalfe, Winget \& Charbonneau (2001)
for the variable WD GD 358.
Variations of the initial mass and chemical composition do not significantly 
affect the main conclusion 
of the present work. In any case, a detailed description of the internal 
chemical stratification 
in WDs generated by progenitors with masses ranging between 1 and 7 M$_\odot$  and
metallicity in the range Z=0 to Z=0.02 has been reported by 
Dom\'\i nguez, H\"oflich \& Straniero (2001).
All the five models were obtained by using the rate provided by Kunz et al. (2002) for the
$^{12}$C$(\alpha,\gamma)^{16}$O reaction. The value of this rate is 
about a factor 2 lower than the one claimed by Metcalfe, Salaris \& Winget (2002) to explain the
observed composition of GD 358. 

Some relevant properties of the computed models are reported in Tab. 1.

\subsection {Bare Schwarzschild Models}
In the case of classical models (Iben \& Rood 1970) the extension of the convective core 
is simply determined by the condition $\nabla_{rad}>\nabla_{ad}$
(the so-called Schwarzschild criterion). No special algorithms to account for the 
induced overshoot and semiconvection are considered. 
We are well aware that this kind of He-burning models is out-of-date.
Nevertheless, they are still used, particularly in studies of stellar progenitors of WDs 
(Althaus et al. 2002) and of SNe Ia (Umeda et al. 1999).

Since bare Schwarzschild criterion only provides a lower limit to the size of the convective core,
the resulting He-burning lifetime is particularly short and  
the final central oxygen abundance 
is definitely lower than the one derived from seismic data of variable WDs
(see Tab. 1).
The evolution of the central composition for this model (BSM - Bare Schwarzschild Model) 
is shown in panel A of Figure 1.
On the other hand, a large fraction of He remains unburned and 
a longer AGB phase takes place. Then, the predicted value
for the  $R_2$ parameter is rather large ($\sim 0.7$), in clear disagreement
with the value obtained for Globular Cluster stars.

\subsection {Semiconvective models}
In this case we have used a numerical algorithm to handle
the induced overshoot and the consequent semiconvection
(see Castellani et al. 1985). Starting from the center of the star, a small mass fraction 
(namely a mesh-point)
is added to the top of the convective core. Then, this procedure  
is iterated until convective 
neutrality is achieved at the external border (convective core) or inside the
well mixed region (convective core plus semiconvective zone). In the latter case, a detached
convective shell forms, whose  
external border is moved outward until 
$\nabla_{rad}=\nabla_{ad}$. The resulting evolution of the internal He profile
is shown in Figure 2 (left panel). Note the ongoing 
growth of the fully convective core (the most internal and flat zone) 
and the formation of the outermost partially
mixed region (semiconvective zone).

As it is well known, when the central He is substantially depleted, 
some instabilities (named breathing pulses) take place at the external border of the convective core.
In this phase, even a small ingestion of fresh He causes a significant increase of the nuclear energy
production, thus leading to a convective runaway.
Basing on both theoretical and observational constraints, breathing
pulses are usually attributed to the adopted algorithms rather than to the physics of convection.
Following the
suggestion of Dorman and Rood (1993), we have suppressed these instabilities 
by setting to zero the gravitational (thermal) energy in the core when the central mass
fraction of He drops below 0.1. 
Note that in this phase, the gravitational energy generally  
accounts for a small positive contribution to the total energy (less than 1\%). Only during a 
breathing pulse,
to counterbalance the sudden increase in the thermonuclear energy flux, 
does the contribution of the gravitational energy become large and negative. 
By neglecting this term in the energy balance equation, 
the external border of the mixed region slowly
recedes and breathing pulses are avoided. This case is reported in Table 1 with the label SM
(semi-convective model).
The corresponding evolutions of the central abundances of He and C are reported in panel
 B (solid lines) of Figure 1.
Note the evident decrease of the He consumption rate in the final
part of the core He-burning phase. This is due to the efficiency of the 
$^{12}$C$(\alpha,\gamma)^{16}$O reaction that
releases almost the same energy as the 3$\alpha$ with 1/3 of fuel consumption. 
This delay of the final part of
the He-burning have a great influence on the final core composition:
a large oxygen
mass fraction (0.79) is left at the center.
This value is only slightly lower than that found for GD 358.
We have already recalled that semiconvevtive models also provide the best reproduction
of the $R_2$ ratio observed in Globular Clusters. 

Obviously, the artificial suppression of the breathing pulses may
be obtained by means of different methods. In principle,
there are no evident reasons to prefer one method with respect to another.  
A rather diffused algorithm has been proposed by Caputo et al. (1989).      
The growth of the convective core is limited by
the constraint that the central He abundance 
can not increase with time, when the central He mass fraction drops below 0.1.
In practice, this method
strongly reduces the effects of the induced overshoot and semiconvection 
during the late stage of core He-burning. As a result,  
the He-exhaustion phase
of this model is similar to the one found in the case of classical models. 
The evolutions of the central mass fractions of He and C for this model are shown in panel B of Figure 1
(dashed lines). This model is reported in Table 1 with the label PSM (partially 
semiconvective models).
As in the case of SM, PSM provides a quite good reproduction of the observed $R_2$ ratio. However, 
since the C depletion essentially takes place during the late stage of the core He-burning,
PSM predicts a moderate oxygen production, very similar to the one obtained in the BSM case.
Note that PSM is the model adopted
by Metcalfe, Winget \& Charbonneau (2001)
in their analysis of the chemical composition of GD 358.

\subsection{Mechanical overshoot}
An alternative scenario could arise if a sizeable mechanical 
overshoot induces an efficient mixing of the region located beyond 
the external border 
of the fully convective core.
In this case, no special algorithms are used to handle semiconvection or breathing pulses  
(see e.g. Girardi et al. 2000)
The case of a moderate overshoot, namely 0.2 $H_p$ (LOM-low overshoot model),
is illustrated in Figure 3, 
where we show the radiative and the adiabatic temperature gradients for three different  
epochs, namely when the residual mass fraction of He at the center is: 0.76, 0.44 and 0.2,
panel A, B and C, respectively. At the beginning of the He-burning (Panel A),  
mechanical overshoot brings the discontinuity of the radiative gradient (solid line)
below the level of the adiabatic gradient (dashed line), so that the external border 
of the convective core is stable. It goes without saying that mechanical overshoot mimics
the job done by the induced overshoot in semiconvective models.
However, as shown in the other two panels, when He-burning goes ahead, 
a convective shell forms, whose external boundary is unstable. A partial mixing beyond 
this boundary (semiconvection) or some form of overshoot,
 capable to restore the convective neutrality, is obviously required.
The resulting evolutions of the central He and C  
are reported in panel C of Figure 1. 
When the central He mass fraction drops below
0.6, some rapid variations of the core composition take place. Their origin is easily understood
by looking at the sequence reported in Figure 3.  Owing to the
overshoot, the size of the convective core increases until its external boundary reaches the
base of the previous convective shell and, in turn, a substantial amount of fresh He is 
suddenly ingested. 

The inconsistency of such kind
of models may be removed if a moderate overshoot is also applied to 
the external border of the convective shell. In this case,
the resulting final composition would be very similar to the  
one obtained in the case of a semiconvective model.  
On the contrary, a significantly different scenario arises if a large mechanical overshoot
takes place at the external border of the convective core.
In this case, the semiconvective zone would be swept away. 
We have calculated an additional model by applying an overshoot of  
1 $H_P$ (HOM-high overshoot model). Under this assumption, the mass of the material within the
classical border is about half of the mass actually mixed. As the He-burning proceeds,
the mixing rapidly extends up to about 0.3 M$_\odot$, so that  
any trace of semiconvection is canceled out (see right panel in Figure 2) and
no convective shells form.  
The evolution of the central composition is shown in panel C of Figure 1 (solid lines). 

It should be noted that overshoot models are not immune to breathing pulses.
For this reason, LOM and HOM have been calculated by adopting the same method used in SM to
avoid these instabilities. A second clarification regards the fact that no overshoot 
has been considered for the computation of the H-burning phase. As it is well known an overshoot
occurring during this phase would produce a larger He core mass, which is equivalent to 
the case of a larger stellar mass without overshoot. However, as we have already reminded,
a change of the stellar mass does not substantially modify the final central amounts of C and O.

In summary, the late stage of the core He-burning in overshoot models is
particularly fast. As a consequence, a smaller amount of oxygen is produced
with respect to fully semiconvective models.
However, if the overshoot is small enough, a semiconvective layer
survives, thus increasing the oxygen production.
For these models, a suitable reproduction of the $R_2$ ratio may be obtained by tuning
the size of the overshoot zone. However this is a calibration of 
the model rather than a prediction. In this framework, a large overshoot is ruled out,
because it would imply too low a value of the $R_2$ parameter (Renzini \& Fusi Pecci, 1988).

\subsection{Time dependent convection}
All the models previously described have been obtained, as usual,
by assuming an instantaneous mixing within the convective core and, eventually, in the overshoot region.
In the last few years some papers report calculations of AGB models obtained by adopting
a time dependent mixing scheme (Herwig et al. 1997; Mazzitelli, D'Antona \& Ventura 1999,
Cristallo et al. 2001, Chieffi et al. 2001).
Note, however, that since the convective velocities are evaluated in the 
framework of the mixing length theory, the resulting mixing within the convective regions 
is very efficient, practically instantaneous. Nevertheless, the relevant point 
for the present work is that during the TP-AGB phase the convective envelope penetrates
the He-rich intershell (the so called third dredge up). Since the opacity at the base of the convective
envelope (H-rich) is larger than that found in the layer located just below (He-rich),
a discontinuity in the radiative gradient takes place. This is exactly the same phenomenon
that occurs at the external border of the convective core during the central He-burning phase.
Indeed, such a discontinuity should induce a further mixing below the formally unstable region,
 namely the one defined by the bare Schwarzschild criterion.
Herwig et al. (1997) propose to add an extra-mixing (or overshoot) outside the boundaries of the formally
convective regions.
In the overshoot zones they assume an exponential decay of the convective velocity, in agreement with 
hydrodynamical simulations of shallow stellar surface convective zone (Freytag et al. 1996). A similar
decay of the convective velocity has been also found by Asida \& Arnett (2000) in 
hydrodynamical calculations of convective regions driven by an oxygen burning
shell in massive stars. Then, in order to check the effect of a time dependent mixing on our result,
we have tentatively computed a model of core He-burning star
by assuming an exponential decay
of the convective velocity beyond the boundaries of the formally convective zones 
(see Chieffi et al. 2001 for details on the numerical algorithm).
 In such a way, a partial mixing is found above the fully convective
core and, eventually, above the convective shell. 
The depth of the partially mixed regions is modulated by the difference between the 
radiative and the adiabatic gradient at the formally convective boundary, but depends
on the adopted strength of the velocity exponential decay, which is free parameter.
In any case,  the overall properties of the
model and, in particular, the final
oxygen abundance are similar to those obtained in the case of semiconvective models,
thus confirming the previous investigation reported by Sweigart (1990).

\section{Discussion and conclusion}
Table 1 summarizes the results of the five models described in the previous section.
Some internal profiles of C and O at the beginning of the thermally pulsing AGB phase are
shown in Figure 4. The innermost region keeps track of the mixing experienced
during core He-burning phase.
In semiconvective models, as well as in the model 
with mechanical overshoot, a sharp variation of the chemical composition 
around M$_r\sim0.3$ M$_\odot$  
separates the region that is partially or totally mixed during the core burning
from the external layers, whose composition remains unchanged\footnote {namely, as fixed
by the first dredge-up} up to the beginning of the double shell burning phase (early-AGB). 
In the case of a large overshoot (HOM - panel C of Figure 4),
 this sharp discontinuity coincides with the maximum
extension attained by the convective core (including the overshoot). Therefore,
its location depends on the assumed value of the overshoot parameter.
In semiconvective models (panels B of Figure 4) two different regions may be
distinguished below the chemical discontinuity: 
a central homogeneous region, created by the fully convective core, 
and an intermediate region, which coincides with the semiconvective zone. 
A bump
in the oxygen distribution characterizes this intermediate region. Owing to the partial mixing
induced by semiconvection, 
a small amount of He still survives in this zone at the epoch of central He exhaustion. 
Therefore, at the beginning of the early-AGB, the temperature rises and the residual
He is rapidly consumed, mainly through the 
$^{12}$C$(\alpha,\gamma)^{16}$O reaction, so that a considerable amount of O is accumulated.
Salaris et al. (1997) suggested that this bump in the oxygen profile is smeared off
by a Rayleigh-Taylor instability.   
In principle, the sharp discontinuity at M$_r\sim0.3$ M$_\odot$ could be
smoothed by the chemical diffusion operating during the long cooling time scale. 
However, only minor effects are expected in the case of bright variable WDs.
In such a case, the location of the chemical discontinuity might be  
derived from seismic data of pulsating WDs. The last column of Table 1 shows
 the expected location of this discontinuity ($M_D$) for semiconvective and overshoot models. 
This may be compared with the value of the $q$ parameter 
reported by Metcalfe, Winget \& Charbonneau (2001) for GD 358. According to these authors, $q$
is the ratio between 
the innermost, almost homogeneous, region and the total mass of the WD.
Metcalfe et al. (2001) report $q=0.49\pm0.01$, which would imply M$_D$=0.29 and 0.32 M$_\odot$ for
M$_{WD}$=0.6 and 0.65 M$_\odot$ respectively. An inspection of Table 1
shows that the calculated values are in very good
agreement with the measured quantity. As noted by Metcalfe, Salaris \& Winget
(2002), since the location of $M_D$ depends on the value of the overshoot parameter, its measure 
could be used to test the efficiency of the mechanical overshoot. We merely comment that this 
may be done only if overshoot is particularly strong ($\sim 1 H_P$ or larger),
otherwise $M_D$ will be in any
case fixed by the extension of the semiconvective region, as discussed in section 2.3.

In the case of the  classical model (panel A of Figure 4), 
the innermost flat region 
coincides with the maximum extension of the convective core during core
He-burning. In this model, the small convective core does not
cover the whole region where nuclear burning is efficient. 
Toward the end of the core burning phase, the He depletion 
takes place well outside the convective core. Later on, at the beginning of the early AGB,
the incoming 
He-burning shell terminates the task initiated during the previous phase. The maximum in the
oxygen profile coincides with the base of the radiative region, where only a small amount
of He survives at the end of the core He burning.  

The second, more stringent, comparison between theory and asteroseismology concerns 
the central oxygen
abundance. The predicted values are in the fourth column of Table 1. The largest oxygen abundance
is obtained in the case of the SM. The calculated mass fraction 
does not differ greatly from 
the value derived from GD 358 by Metcalfe et al. (2001) and is probably well within the error bars 
of the seismic data. 
This result has been obtained by using a moderate rate for
the $^{12}$C$(\alpha,\gamma)^{16}$O reaction, as
recently reported by Kunz et al. (2002). All the other models lead to 
lower abundances for the central oxygen. 
As already discussed by Metcalfe et al. (2002), a very large rate of the carbon destruction 
would be required to reconcile these models with the measured central oxygen abundance in GD 358 . 

The evaluation of the experimental error affecting the measurements
of the $^{12}$C$(\alpha,\gamma)^{16}$O reaction rate is a hard task. 
A complete bibliography of the published experimental
reports may be found in the paper of Kunz et al. (2002).
The laboratory experiments have been  extended down to
about 1 MeV. Below this energy, the extremely small
value of the cross section ($<10$ pb) hampers direct detection of $\gamma$-rays and 
extrapolation procedures have to be used to extract the astrophysical 
S-factor at the energy relevant for core He-burning ($\sim$300 keV).
Such an extrapolation is based on
the fitting of differential cross sections in the investigated region and 
requires the inclusion of the phase correlation between the two incoming partial
waves that contribute to the two multipoles. 
A global analysis that attempts to take into account all the possible sources of uncertainty 
has been reported by 
Buchmann (1996, 1997). According to this work, at $T=1.8 \cdot 10^8$ K (about 300 keV), 
the possible value of the reaction rate ranges between 
$N_A<\sigma,v>=0.5$ and 2.2 ($10^{-15} cm^3/mol/s$). 
A recent compilation by NACRE (Angulo et al. 1999) adopts a slightly smaller range of
accepted values, namely between 0.9 
and 2.1  ($10^{-15} cm^3/mol/s$).
Finally, the latest experimental investigation by Kunz et al. (2002) 
reports $N_A<\sigma,v>=1.25$ ($10^{-15} cm^3/mol/s$) $\pm 30\%$. 

We have investigated the influence of this uncertainty on our conclusions. Some additional 
models have been obtained by multiplying the 
rate of the $^{12}$C$(\alpha,\gamma)^{16}$O by a factor of f=0.4 and 1.6. This corresponds
to an error bar which is double the one quoted by Kunz et al. (2002) and similar to the one
accepted by Buchmann (1996). The resulting central
oxygen abundances for SM and HOM are reported in the last four rows of Table 1.
Even by considering such a large error, the maximum amount of central oxygen in models
without semiconvection is definitely lower than that claimed by Metcalfe et al. (2001) for GD 358.

In summary, by combining the uncertainty on the relevant nuclear reaction rate and that due to
convective efficiency, the theoretical predictions on the final central oxygen mass fraction 
range between 0.3 and 0.9. The large value reported by Metcalfe et al.  
favors semiconvective models. However, a moderate mechanical overshoot, which leaves unchanged
the semiconvective layer, is still possible.
     
\acknowledgments
We are grateful to T. Metcalfe and G. Fontaine for their enlightening illustrations on the
potential of the asteroseismology applied to WDs and to F. Terrasi and L. Gialanella
for their contributions to our understanding of the latest developments in nuclear physics
experiments. This work has been partially supported by the Italian grant MURST-Cofin2001 
 and by the Spanish grants AYA2000-1574 and 
 FQM-292.


\clearpage

\begin{figure}
\plotone{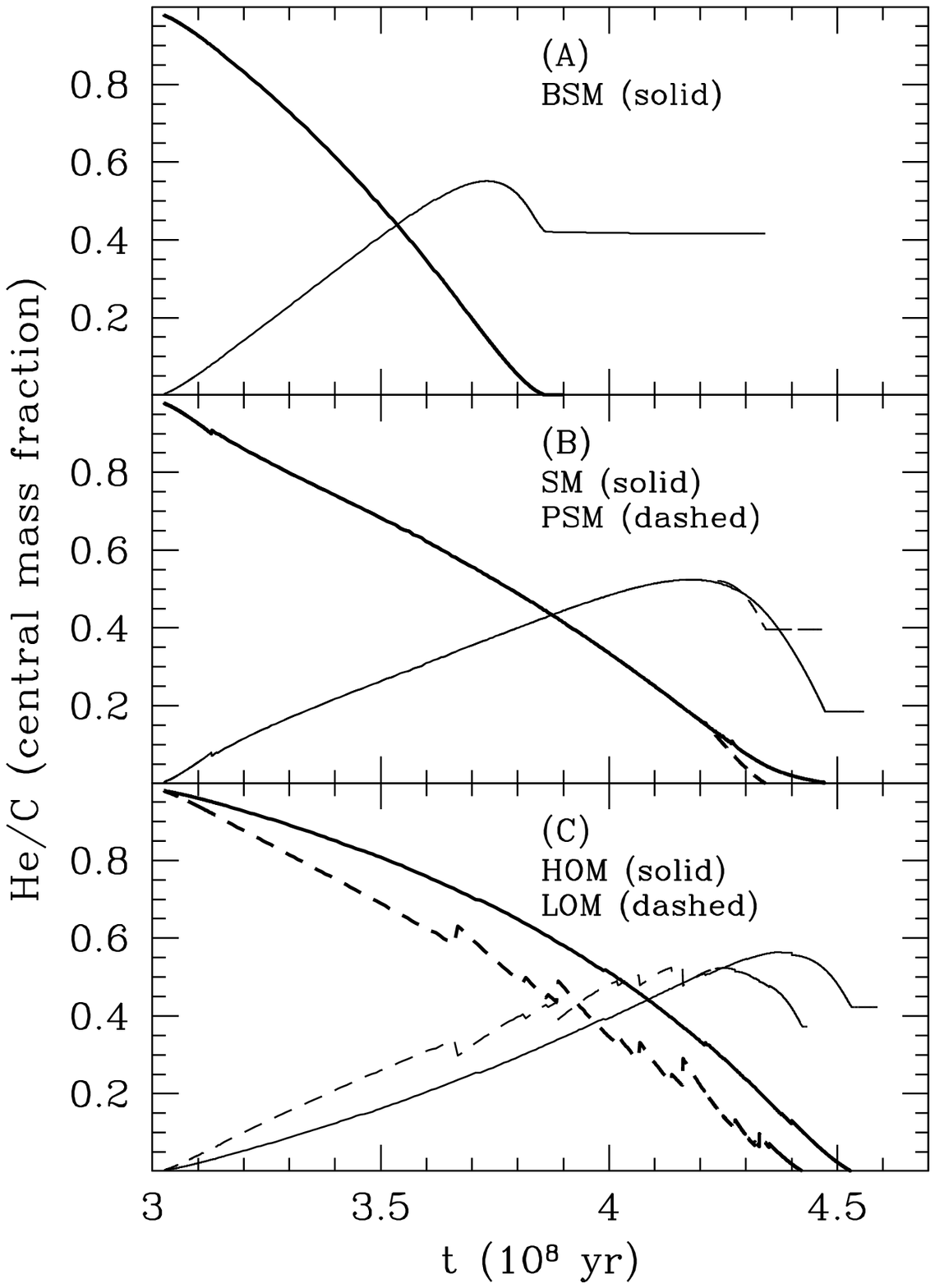}
\caption{Evolution of the central He mass fraction (thick line) and that of the
central C mass fraction (thin line) as obtained by changing the treatment of the
core convection: BSM -  bare Schwarzschild model (no overshoot, no semiconvection),
SM - semiconvective model (breathing pulse suppression
as in Dorman \& Rood 1993), PSM - semiconvective model (breathing pulse suppression as
in Caputo et al. 1989), HOM - high overshoot model (1 $H_p$), LOM - low overshoot model (0.2 $H_p$).
\label{fig1}}
\end{figure}

\begin{figure}
\plotone{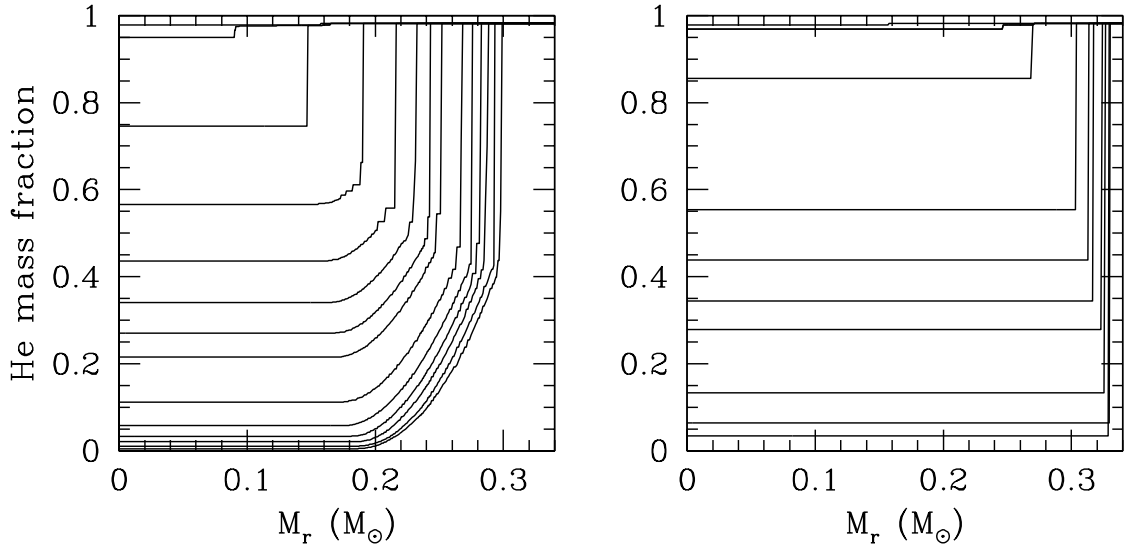}
\caption{Evolution of the internal He profile for the semiconvective model (SM, left panel) and
for the high overshooting model (HOM, right panel). \label{fig2}}
\end{figure}

\begin{figure}
\plotone{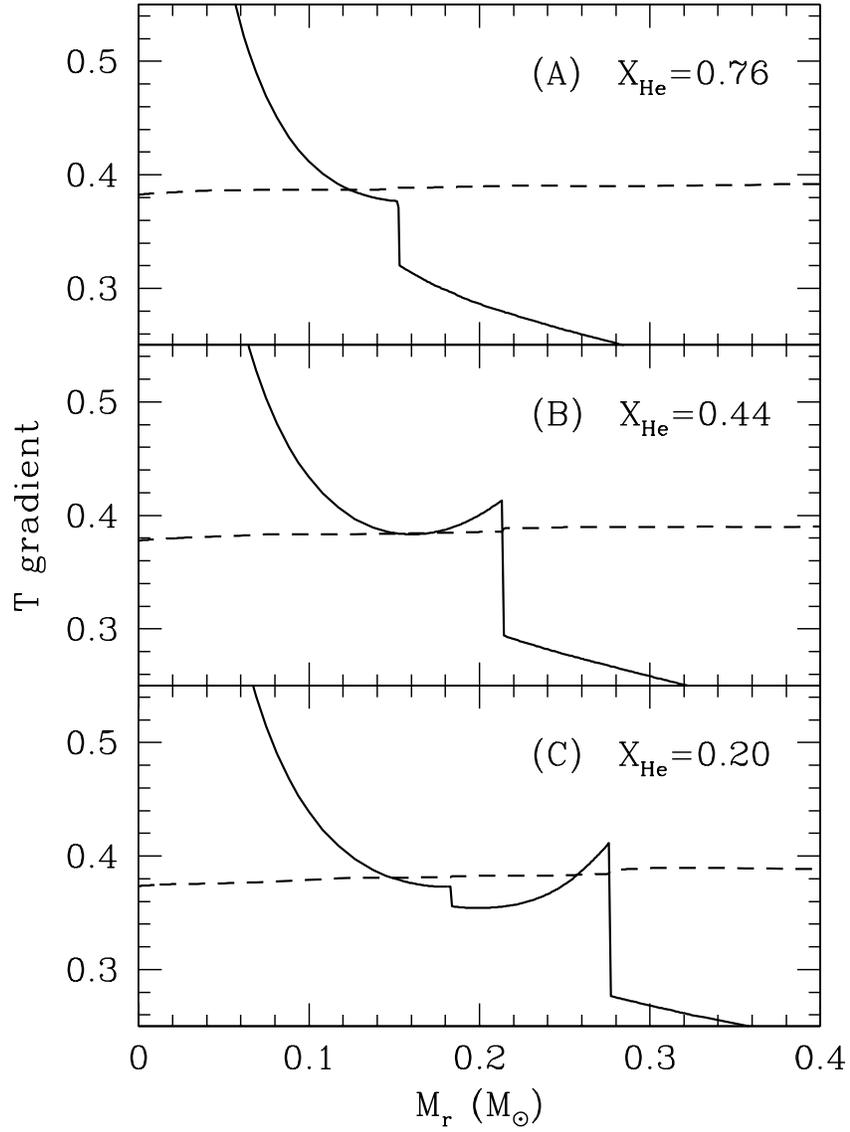}
\caption{Radiative (solid) and adiabatic (dashed) temperature gradients in the core
of the LOM model (0.2 $H_p$ overshoot model). X$_{He}$ is the central
mass fraction of He. \label{fig3}}
\end{figure}

\begin{figure}
\plotone{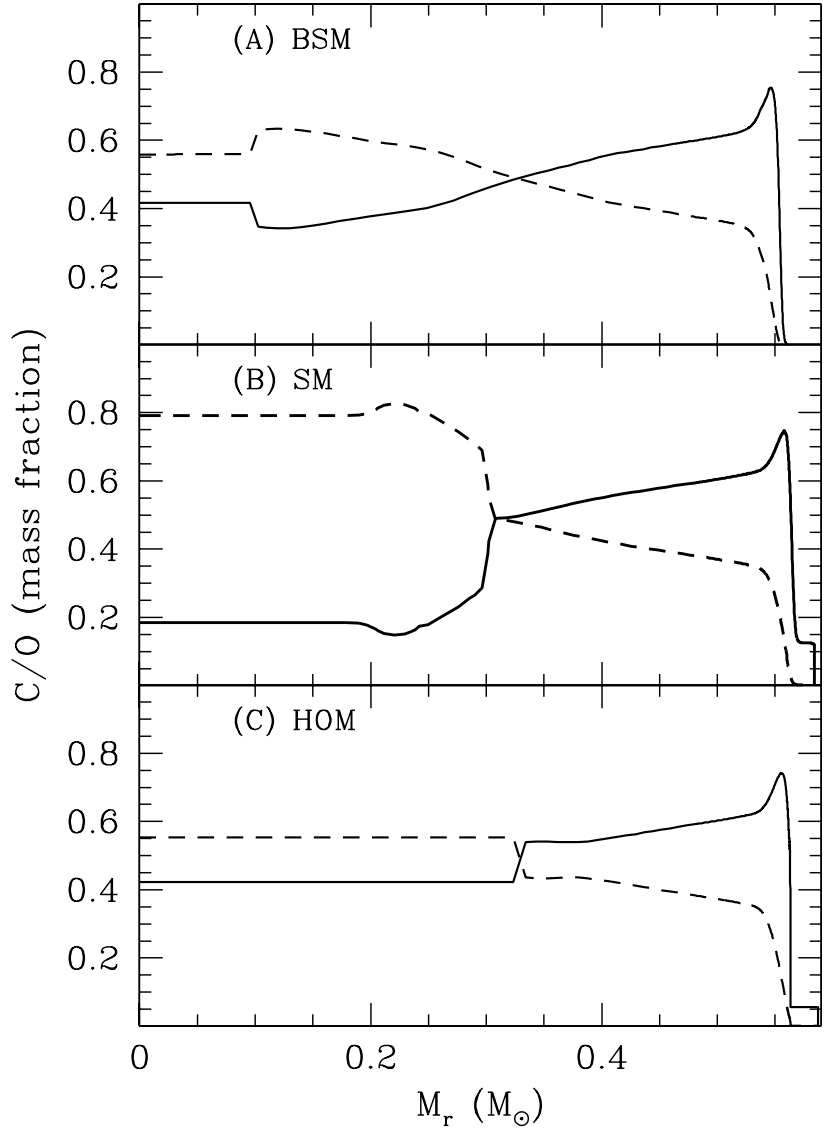}
\caption{Chemical stratification, C (solid) and O (dashed), of the core at
 the onset of the thermally pulsing AGB phase. \label{fig4}}
\end{figure}


\clearpage

\begin{deluxetable}{cccccc} 
\tablecaption{Core He-burning Models: M=3 M$_\odot$ Z=0.02 \label{tab1}}
\tablewidth{0pt}
\tablehead {
\colhead {label} &
\colhead {f \tablenotemark{1}} &
\colhead {$\tau_{He}$\tablenotemark{2}} & 
\colhead {$X_C$\tablenotemark{3}} &
\colhead {$X_O$\tablenotemark{3}} &
\colhead {M$_D$\tablenotemark{4}} } 

\startdata
BSM       &   1  &  88  &  0.42  & 0.56 &       \\
SM        &   1  &  145 &  0.19  & 0.79 & 0.31  \\  
PSM       &   1  &  134 &  0.40  & 0.58 & 0.27  \\
HOM       &   1  &  153 &  0.42  & 0.56 & 0.32  \\
LOM       &   1  &  139 &  0.38  & 0.60 & 0.28  \\
SM        &  0.4 &  135 &  0.52  & 0.46 & 0.29  \\  
SM        &  1.6 &  149 &  0.08  & 0.90 & 0.31  \\  
HOM       &  0.4 &  142 &  0.66  & 0.32 & 0.31  \\
HOM       &  1.6 &  157 &  0.28  & 0.70 & 0.32  \\
\enddata

\tablenotetext{1}{enhancement factor of the $^{12}$C$(\alpha,\gamma)^{16}$O rate. f=1 corresponds
 to the Kunz et al. (2002) rate.}
\tablenotetext{2}{He-burning lifetime (Myr).}
\tablenotetext{3}{final central mass fractions of C and O.}
\tablenotetext{4}{location, in M$_\odot$, of the sharp discontinuity that marks
the separation between the innermost low C region and the external zone unchanged by the
central convective episodes experienced by the star during core He-burning phase (see Figure 4)} 
\end{deluxetable}


\newpage
\begin{thebibliography}{99}
\bibitem[]{AA} Asida, S.M., \& Arnet, D., 2000, \apj, 545, 435.
\bibitem[]{ASC} Althaus, L.G., Serenelli, A.M., C\'orsico, A.H., \& Benvenuto, O.G., 2002, \mnras, 330, 685.
\bibitem[]{Na} Angulo, C. {\em et al.}, 1999, Nucl. Phys. A, 656, 3
\bibitem[]{bbc} Bertelli, G., Bressan, A., \& Chiosi, C., 1985, \aap, 13, 279.
\bibitem[]{ber} Bertelli, G., Betto, R., Chiosi, C., Bressan, A., \& Nasi, E., 1990, \aaps, 85, 845.
\bibitem[]{bbb} Bressan, A., Fagotto, F., Bertelli, G., \& Chiosi, C., 1993, \aaps, 100, 647.
\bibitem[]{B96} Buchmann L. 1996, \apj, 468, 127.
\bibitem[]{B97} Buchmann L. 1997, \apj, 479, 153.
\bibitem[]{BCFP85} Buonanno R., Fusi Pecci F., \& Corsi C.E., 1985, \aap, 145, 97.
\bibitem[]{BFPBC83} Buzzoni A., Fusi Pecci F., Buonanno R., \& Corsi C.E., 1983, \aap, 128, 94.
\bibitem[]{C89} Caputo F., Castellani, V., Chieffi, A., Pulone, L., \& Tornamb\'e, A., 1989, \apj, 340, 241.
\bibitem[]{CGR71a} Castellani, V., Giannone P., \& Renzini A., 1971a, Ap\&SS, 10, 340.
\bibitem[]{CGR71b} Castellani, V., Giannone P., \& Renzini A., 1971b, Ap\&SS, 10, 355.
\bibitem[]{CCPT85} Castellani, V., Chieffi, A, Pulone, L., \& Tornamb\`e, A., 1985, \apj, 296, 204. 
\bibitem[]{Cetal} Chieffi, A., Dom\'\i nguez, I., Limongi, M., \& Straniero, O., 2001, \apj, 554, 1159. 
\bibitem[]{Cristallo} Cristallo, S., Straniero, O., Gallino, R., Herwig, F., Chieffi, A., Limongi, M.,
         \& Busso, M., 2001, Nucl. Phys. A, 688, 217.  
\bibitem[]{DM} Demarque, P., \& Mengel, J.G., 1972, \apj, 171, 583.
\bibitem[]{DSG} Demarque, P., Sarajedini, A., \& Guo, X.-J., 1994, \apj, 426, 165.
\bibitem[]{DHS01} Dom\'\i nguez, I., H\"oflich, P., \& Straniero, O., 2001, \apj, 557, 279.
\bibitem[]{DR93} Dorman, B., \& Rood, R.T., 1993, \apj, 409, 387.
\bibitem[]{FR} Freytag, B., Ludwig, H.-G., \& Steffen, M., 1996 \aap, 313, 497. 
\bibitem[]{GIR} Girardi, L., Bressan, A., Bertelli, G., \& Chiosi, C., 2000, \aaps, 141, 371.
\bibitem[]{h97} Herwig, F., Bloecker, T., Schoenberner, D., \& El Eid, M., 1997, \aap, 324, L81.
\bibitem[]{I85} Iben, I., 1986, \apj, 304, 201.
\bibitem[]{IR70} Iben, I., \& Rood, R.T., 1970, \apj, 161, 587.
\bibitem[]{I01} Imbriani, G., Limongi, M., Gialanella, L., Straniero, O., \&  Chieffi, A., 2001, \apj,  558, 903.
\bibitem[]{K02} Kunz, R., Fey, M., Jaeger, M., Mayer, A., Hammer, J.W., Staudt, G., Harissopulos, S., \& Paradellis, T., 2002 \apj, 567, 643.
\bibitem[]{L91} Lattanzio, J., 1991, \apjs, 78, 215.
\bibitem[]{Ma75} Maeder, A., 1975, \aap, 43, 61.
\bibitem[]{MDV} Mazzitelli, I., D'Antona, F., \& Ventura, P., 1999, \aap, 348, 845
\bibitem[]{MMM} Mermilliod, J.-C., Huestamendia, G., \& del Rio, G., 1994, \aaps, 106, 419.
\bibitem[]{MNW00} Metcalfe, T.S., Nather, R.E., \& Winget, D.E., 2000, \apj, 545, 974.
\bibitem[]{MWC01} Metcalfe, T.S., Winget, D.E., \& Charbonneau, P., 2001, \apj, 557, 1021.
\bibitem[]{MSW02} Metcalfe, T.S., Salaris, M., Winget, D.E., 2002, \apj, 573, 803. 
\bibitem[]{PT01} Paczi\'nsky, B., 1970a, Acta Astron., 20,47.
\bibitem[]{PT02} Paczi\'nsky, B., 1970b, Acta Astron., 20,195.
\bibitem[]{RF88} Renzini, A., \& Fusi Pecci, F., 1988, \araa, 26, 199.
\bibitem[]{R87} Renzini, A., 1987, \aap, 188, 49.
\bibitem[]{RF72} Robertson, J.W., \& Faulkner, D.J., 1972, \apj, 171, 309.
\bibitem[]{sa97} Salaris, M., Dominguez, I., Garcia-Berro, E., Hernanz, M., Isern,J., \& Mochkovitch, R., 1997, \apj, 486, 413
\bibitem[]{SS65} Saslaw, W.C., \& Schwarzschild, M., 1965, \apj, 142, 1468.
\bibitem[]{Sch} Schaller, G., Schaerer, D., Meynet, G., \& Maeder, A., 1992, \aaps, 96, 269.
\bibitem[]{SPE} Schr\"oder, K-P., Pols, O.R., \& Eggleton, P.P., 1997, \mnras, 285, 696.
\bibitem[]{ss73} Shaviv, G., \& Salpeter, E.E., 1973, \apj, 184, 91.
\bibitem[]{st92} Stothers, R.B., \& Chin, C.-W., 1992, \apj, 390, 136.
\bibitem[]{SD72} Sweigart, A.V., \& Demarque, P., 1972, \aap, 20, 445.
\bibitem[]{SG78} Sweigart, A.V., \& Gross, P.G., 1976, \apjs, 32, 367. 
\bibitem[]{S90} Sweigart, A.V., 1990, Proceedings of the Conference {\em Confrontation between
        pulsation and evolution}, ASP conf. ser., 11, 1.
\bibitem[]{t99} Testa, V., Ferraro, F.R., Chieffi, A., Straniero, O., Limongi, M., \&
        Fusi Pecci, F., 1999, \aj, 118, 2839.
\bibitem[]{u99} Umeda, H., Nomoto, K., Yamaoka, H., \& Wanajo, S., 1999, \apj, 513, 861.
\bibitem[]{vh} Van Horn, H.M., 1971, Proceedings of the conference {\em White Dwarfs}, 
        IAU symp., 42, 97.

\end{thebibliography}
\end{document}